# Enhancement of Cavity Cooling of a Micromechanical Mirror Using Parametric Interactions

Sumei Huang and G. S. Agarwal

*Department of Physics, Oklahoma State University, Stillwater, Oklahoma 74078, USA*

(9th Oct 2008)

It is shown that an optical parametric amplifier inside a cavity can considerably improve the cooling of the micromechanical mirror by radiation pressure. The micromechanical mirror can be cooled from room temperature 300 K to sub-Kelvin temperatures, which is much lower than what is achievable in the absence of the parametric amplifier. Further if in case of a precooled mirror one can reach millikelvin temperatures starting with about 1 K. Our work demonstrates the fundamental dependence of radiation pressure effects on photon statistics.

PACS numbers: 42.50.Lc, 03.65.Ta, 05.40.-a

## I. INTRODUCTION

Recently there is considerable interest in micromechanical mirrors. These are macroscopic quantum mechanical systems and the important question is how to reach their quantum characteristics [1-4]. The thermal noise limits many highly sensitive optical measurements [5,6]. We also note that there has been considerable interest in using micromirrors for producing superpositions of macroscopic quantum states if such micromirrors can be cooled to their quantum ground states [7,8]. Thus cooling of micromechanical resonators becomes a necessary prerequisite for all such studies. So far two different ways to cool a mechanical resonator mode have been proposed. One is active feedback scheme [9-12], where a viscous force is fed back to the movable mirror to decrease its Brownian motion.

The other is passive feedback scheme [4,13-17], in which the Brownian motion of the movable mirror is damped by the radiation pressure force exerted by photons in an appropriately detuned optical cavity.

Clearly we need to think of methods which can cool the micromirror toward its ground state. Since radiation pressure depends on the number of photons, one would think that the cooling of the micromirror can be manipulated by using effects of the photon statistics. In this paper, we propose and analyze a new method to achieve cooling of a movable mirror to sub-Kelvin temperatures by using a type I optical parametric amplifier inside a cavity. We remind the reader of the great success of cavities with parametric amplifiers in the production of nonclassical light [18-20]. The movable mirror can reach a minimum temperature of about few hundred mK, a factor of 500 below room temperature 300K. The lowering of the temperature is achieved by changes in photon statistics due to parametric interactions [21-26]. Note that if the mirror is already precooled to say about 1 K, then we show that by using OPA we can cool to about millikelvin temperatures or less.

The paper is organized as follows. In section II we describe the model and derive the quantum Langevin equations. In section III we obtain the stability conditions, calculate the spectrum of fluctuations in position and momentum of the movable mirror, and define the effective temperature of the movable mirror. In section IV we show how the movable mirror can be effectively cooled by using parametric amplifier inside the cavity.

## II. MODEL

We consider a degenerate optical parametric amplifier (OPA) inside a Fabry-Perot cavity with one fixed partially transmitting mirror and one movable totally reflecting mirror in contact with a thermal bath in equilibrium at temperature $T$, as shown in the Fig. 1. The movable mirror is free to move along the cavity axis and is treated as a quantum mechanical harmonic oscillator with effective mass $m$, frequency $\omega_m$ and energy decay rate $\gamma_m$. The effect of the thermal bath can be modeled by a Langevin force. The cavity field is driven by an input laser field with frequency $\omega_L$ and positive amplitude related to the input laser power $P$ by $\tilde{\varepsilon} = \sqrt{P/(\hbar\omega_L)}$. When photons in the cavity reflect off the surface of the movable mirror, the movable mirror will receive the action of the radiation pressure force, which is proportional to the instantaneous photon number inside the cavity. So the mirror can oscillate under the effects of the thermal Langevin force and the radiation pressure force. Meanwhile, the movable mirror's motion changes the length of the cavity; hence the movable mirror displacement from its equilibrium position will induce a phase shift on the cavity field.

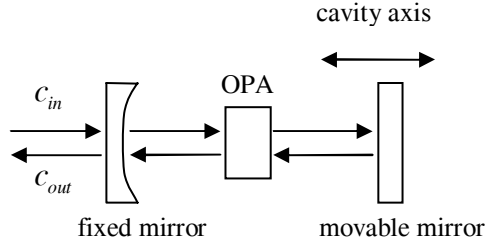

FIG. 1. Sketch of the cavity used to cool micromechanical mirror. The cavity contains a nonlinear crystal which is pumped by a laser (not shown) to produce parametric amplification and to change photon statistics in the cavity.

Here we assume the system is in the adiabatic limit, which means $\omega_m \ll \pi c/L$; $c$ is the speed of light in vacuum and $L$ is the cavity length in the absence of the cavity field. We assume that the motion of the mirror is so slow that the scattering of photons to other cavity modes can be ignored, thus we can consider one cavity mode only [27,28], say, $\omega_c$. Moreover, in the adiabatic limit, the number of photons generated by the Casimir effect [29], retardation and Doppler effects is negligible [9,30,31]. Under these conditions, the total Hamiltonian for the system in a frame rotating at the laser frequency $\omega_L$ can be written as

$$H = \hbar(\omega_c - \omega_L)n_c - \hbar\chi n_c q + \frac{1}{2}(\frac{p^2}{m} + m\omega_m^2 q^2) + i\hbar\varepsilon(c^+ - c) + i\hbar G(e^{i\theta}c^{+2} - e^{-i\theta}c^2). \quad (1)$$

Here $c$ and $c^+$ are the annihilation and creation operators for the field inside the cavity, respectively; $n_c = c^+c$ is the number of the photons inside the cavity; $q$ and $p$ are the position and momentum operators for the movable mirror. The parameter $\chi = \omega_c/L$ is the coupling constant between the cavity and the movable mirror; and $\varepsilon = \sqrt{2\kappa}\tilde{\varepsilon}$. Note that $\kappa$ is the photon decay rate due to the photon leakage through the fixed partially transmitting mirror. Further $\kappa = \pi c/(2FL)$, where $F$ is the cavity finesse. In Eq. (1), $G$ is the nonlinear gain of the OPA, and $\theta$ is the phase of the field driving the OPA. The parameter $G$ is proportional to the pump driving the OPA.

In Eq. (1), the first term corresponds to the energy of the cavity field, the second term arises from the coupling of the movable mirror to the cavity field via radiation pressure, the third term gives the energy of the movable mirror, the fourth term describes the coupling between the input laser field and the cavity field, the last term is the coupling between the OPA and the cavity field.

The motion of the system can be described by the Heisenberg equations of motion and adding the corresponding damping and noise terms, which leads to the following quantum Langevin equations

$$\dot{q} = p/m,$$

$$\dot{p} = -m\omega_m^2 q + \hbar\chi n_c - \gamma_m p + \xi,$$

$$\dot{c} = i(\omega_L - \omega_c)c + i\chi qc + \varepsilon + 2Ge^{i\theta}c^+ - \kappa c + \sqrt{2\kappa}c_{in}. \quad (2)$$

Here $c_{in}$ is the input vacuum noise operator with zero mean value; its correlation function is [32]

$$\langle \delta c_{in}(t)\delta c_{in}^+(t')\rangle = \delta(t-t'),$$

$$\langle \delta c_{in}(t)\delta c_{in}(t')\rangle = \langle \delta c_{in}^+(t)\delta c_{in}(t')\rangle = 0. \quad (3)$$

The force $\xi$ is the Brownian noise operator resulting from the coupling of the movable mirror to the thermal bath, whose mean value is zero, and it has the following correlation function at temperature $T$ [31]

$$\langle \xi(t)\xi(t')\rangle = \frac{\hbar\gamma_m}{2\pi}m\int \omega e^{-i\omega(t-t')}[\coth(\frac{\hbar\omega}{2k_B T})+1]d\omega, \quad (4)$$

where $k_B$ is the Boltzmann constant and $T$ is the thermal bath temperature. In order to analyze Eq. (2), we use standard methods from Quantum Optics [33]. A detailed calculation of the temperature for $G = 0$ is given by Paternostro *et al.* [16]. By setting all the time derivatives in Eq. (2) be zero, we obtain the steady-state mean values

$$p_s = 0, \quad q_s = \frac{\hbar\chi|c_s|^2}{m\omega_m^2}, \quad c_s = \frac{\kappa - i\Delta + 2Ge^{i\theta}}{\kappa^2 + \Delta^2 - 4G^2}\varepsilon, \quad (5)$$

where

$$\Delta = \omega_c - \omega_L - \chi q_s = \Delta_0 - \chi q_s = \Delta_0 - \frac{\hbar\chi^2|c_s|^2}{m\omega_m^2} \quad (6)$$

is the effective cavity detuning, including the radiation pressure effects. The modification of the detuning by the $\chi q_s$ term depends on the range of parameters. The $q_s$ denotes the new equilibrium position of the movable mirror relative to that without the driving field. Further $c_s$ represents the steady-state amplitude of the cavity field. Note that $q_s$ and $c_s$ can display optical multistable behavior, which is a nonlinear effect induced by the radiation-pressure coupling of the movable mirror to the cavity field. Mathematically this is contained in the dependence of the detuning parameter $\Delta$ on the mirror's amplitude $q_s$. It is evident from Eqs. (5) and (6) that $\Delta$ satisfies a fifth order equation and in principle can have 5 real solutions implying multistability. Generally, in this case, at most three solutions would be stable. The bistable behavior is reported in Refs [34,35].

### III. RADIATION PRESSURE AND QUANTUM FLUCTUATIONS

In order to determine the cooling of the mirror, we need to find out the fluctuations in the mirror's amplitude. Since the problem is nonlinear, we assume that the nonlinearity is weak. We are thus interested in the dynamics of small fluctuations around the steady state of the system. Such a linearized analysis is quite common in quantum optics [33,36]. So we write each operator of the system as the sum of its steady-state mean value and a small fluctuation with zero mean value,

$$q = q_s + \delta q, p = p_s + \delta p, c = c_s + \delta c . \tag{7}$$

Inserting Eq. (7) into Eq. (2), then assuming $|c_s| \gg 1$, we get the linearized quantum Langevin equations for the fluctuation operators

$$\delta \dot{q} = \delta p / m ,$$

$$\delta \dot{p} = -m\omega_m^2 \delta q + \hbar \chi (c_s \delta c^+ + c_s^* \delta c) - \gamma_m \delta p + \xi ,$$

$$\delta \dot{c} = -i\Delta \delta c + i\chi c_s \delta q + 2Ge^{i\theta} \delta c^+ - \kappa \delta c + \sqrt{2\kappa} \delta c_{in} ,$$

$$\delta \dot{c}^+ = i\Delta \delta c^+ - i\chi c_s^* \delta q + 2Ge^{-i\theta} \delta c - \kappa \delta c^+ + \sqrt{2\kappa} \delta c_{in}^+ . \tag{8}$$

Introducing the cavity field quadratures $\delta x = \delta c^+ + \delta c$ and $\delta y = i(\delta c^+ - \delta c)$, and the input noise quadratures $\delta x_{in} = \delta c_{in}^+ + \delta c_{in}$ and $\delta y_{in} = i(\delta c_{in}^+ - \delta c_{in})$, Eq. (8) can be written in the matrix form

$$\dot{f}(t) = Af(t) + \eta(t) , \tag{9}$$

where $f(t)$ is the column vector of the fluctuations, $\eta(t)$ is the column vector of the noise sources. For the sake of simplicity, their transposes are

$$f(t)^T = (\delta q, \delta p, \delta x, \delta y) ,$$

$$\eta(t)^T = (0, \xi, \sqrt{2\kappa}\delta x_{in}, \sqrt{2\kappa}\delta y_{in}) ; \tag{10}$$

and the matrix $A$ is given by

$$A = \begin{pmatrix} 0 & \frac{1}{m} & 0 & 0 \\ -m\omega_m^2 & -\gamma_m & \hbar\chi\frac{c_s + c_s^*}{2} & \hbar\chi\frac{c_s - c_s^*}{2i} \\ i\chi(c_s - c_s^*) & 0 & -(\kappa - 2G\cos\theta) & \Delta + 2G\sin\theta \\ \chi(c_s + c_s^*) & 0 & -\Delta + 2G\sin\theta & -(\kappa + 2G\cos\theta) \end{pmatrix}. \tag{11}$$

The solutions to Eq. (9) are stable only if all the eigenvalues of the matrix $A$ have negative real parts. Applying the Routh-Hurwitz criterion [37,38], we get the stability conditions

$$2\kappa(\kappa^2 - 4G^2 + \Delta^2 + 2\kappa\gamma_m) + \gamma_m(2\kappa\gamma_m + \omega_m^2) > 0,$$

$$(2\kappa + \gamma_m)^2 [\frac{2\hbar\chi^2 |c_s|^2}{m}\Delta + \frac{2\hbar\chi^2(c_s^2 + c_s^{*2})G\sin\theta}{m} + \frac{2i\hbar\chi^2(c_s^2 - c_s^{*2})G\cos\theta}{m}] + 2\kappa\gamma_m\{(\kappa^2 - 4G^2 + \Delta^2)^2$$

$$+ (2\kappa\gamma_m + \gamma_m^2)(\kappa^2 - 4G^2 + \Delta^2) + \omega_m^2 [2(\kappa^2 + 4G^2 - \Delta^2) + \omega_m^2 + 2\kappa\gamma_m]\} > 0,$$

$$\omega_m^2(\kappa^2 - 4G^2 + \Delta^2) - \frac{2\hbar\chi^2 |c_s|^2}{m}\Delta - \frac{2\hbar\chi^2(c_s^2 + c_s^{*2})G\sin\theta}{m} - \frac{2i\hbar\chi^2(c_s^2 - c_s^{*2})G\cos\theta}{m} > 0 . \tag{12}$$

Note that in the absence of coupling $\chi$, the conditions (12) become equivalent to

$$\kappa^2 - 4G^2 + \Delta^2 > 0. \tag{13}$$

The condition for the threshold for parametric oscillations is $\kappa^2 - 4G^2 + \Delta^2 = 0$. We always would work under the condition that (13) is satisfied. Further for $\chi \neq 0$ we would do numerical simulations using parameters so that conditions (12) are satisfied.

On Fourier transforming all operators and noise sources in Eq. (8) and solving it in the frequency domain, the position fluctuations of the movable mirror are given by

$$\delta q(\omega) = -\frac{1}{d(\omega)}\{[\Delta^2 + (\kappa - i\omega)^2 - 4G^2]\xi(\omega) - i\hbar\sqrt{2\kappa}\chi[((\omega + i\kappa - \Delta)c_s + 2iGe^{i\theta}c_s^*)\delta c_{in}^+(\omega)$$

$$+ ((\omega + i\kappa + \Delta)c_s^* + 2iGe^{-i\theta}c_s)\delta c_{in}(\omega)]\}, \tag{14}$$

Where $d(\omega) = 2\hbar\chi^2(\Delta|c_s|^2 + iGe^{-i\theta}c_s^2 - iGe^{i\theta}c_s^{*2}) + m(\omega^2 - \omega_m^2 + i\omega\gamma_m)[\Delta^2 + (\kappa - i\omega)^2 - 4G^2]$. In Eq. (14), the first term proportional to $\xi(\omega)$ originates from the thermal noise, while the second term proportional to $\chi$ arises from radiation pressure. So the position fluctuations of the movable mirror are now determined by the thermal noise and

radiation pressure. Notice that if there is no radiation pressure, the movable mirror will make Brownian motion, $\delta q(\omega) = -\xi(\omega)/[m(\omega^2 - \omega_m^2 + i\omega\gamma_m)]$, whose susceptibility has a Lorentzian shape centered at frequency $\omega_m$ with width $\gamma_m$.

The spectrum of fluctuations in position of the movable mirror is defined by

$$S_q(\omega) = \frac{1}{4\pi} \int d\Omega e^{-i(\omega+\Omega)t} \langle \delta q(\omega)\delta q(\Omega) + \delta q(\Omega)\delta q(\omega) \rangle. \tag{15}$$

To calculate the spectrum, we need the correlation functions of the noise sources in the frequency domain,

$$\langle \delta c_{in}(\omega)\delta c_{in}^+(\Omega) \rangle = 2\pi\delta(\omega+\Omega),$$

$$\langle \xi(\omega)\xi(\Omega) \rangle = 2\pi\hbar\gamma_m m\omega[1 + \coth(\frac{\hbar\omega}{2k_B T})]\delta(\omega+\Omega). \tag{16}$$

Substituting Eq. (14) and Eq. (16) into Eq. (15), we obtain the spectrum of fluctuations in position of the movable mirror

$$S_q(\omega) = \frac{\hbar}{|d(\omega)|^2} \{ 2\kappa\hbar\chi^2[(\kappa^2+\omega^2+\Delta^2+4G^2)|c_s|^2 + 2Ge^{i\theta}c_s^{*2}(\kappa-i\Delta) + 2Ge^{-i\theta}c_s^2(\kappa+i\Delta)]$$

$$+ m\gamma_m\omega[(\Delta^2+\kappa^2-\omega^2-4G^2)^2 + 4\kappa^2\omega^2] \times \coth(\frac{\hbar\omega}{2k_B T}) \}. \tag{17}$$

In Eq. (17), the first term is the radiation pressure contribution, whereas the second term corresponds to the thermal noise contribution. Then Fourier transforming $\delta \dot{q} = \delta p/m$ in Eq. (8), we get $\delta p(\omega) = -im\omega\delta q(\omega)$, which leads to the spectrum of fluctuations in momentum of the movable mirror

$$S_p(\omega) = m^2\omega^2 S_q(\omega). \tag{18}$$

For a system in thermal equilibrium, we can use the equipartition theorem to define temperature $\frac{1}{2}m\omega_m^2\langle q^2 \rangle = \frac{\langle p^2 \rangle}{2m} = \frac{1}{2}k_B T_{eff}$, where $\langle q^2 \rangle = \frac{1}{2\pi}\int_{-\infty}^{+\infty} S_q(\omega)d\omega$, and $\langle p^2 \rangle = \frac{1}{2\pi}\int_{-\infty}^{+\infty} S_p(\omega)d\omega$. However here we are dealing with a driven system and $\frac{1}{2}m\omega_m^2\langle q^2 \rangle \neq \frac{\langle p^2 \rangle}{2m}$, hence the question is how to define temperature. We use an effective temperature defined by the total energy of the movable mirror $k_B T_{eff} = \frac{1}{2}m\omega_m^2\langle q^2 \rangle + \frac{\langle p^2 \rangle}{2m}$. We also introduce the parameter $r = m^2\omega_m^2\langle q^2 \rangle / \langle p^2 \rangle$. This parameter gives us the relative importance of fluctuations in position and momentum of the mirror. We mention that one can calculate the quantum state of the oscillator and we find that the Wigner function is Gaussian.

The Eq. (17) is our key result which tells how the temperature of the micromirror would depend on the parameters of the cavity: $\kappa$, gain of the OPA, external laser power etc. We specifically investigate the dependence of the temperature on the gain $G$ and the phase $\theta$ associated with the parametric amplification process. In the limit of $G \to 0$, the result (17) reduces to the one derived by Paternostro *et al.* [16].

## IV. COOLING MIRROR TO ABOUT SUB-KELVIN TEMPERATURES

In this section, we present the possibility of cooling the micromirror to temperatures of about sub-Kelvin by using parametric amplifiers inside cavities. In all the numerical calculations we choose the values of the parameters which are similar to those used in recent experiments: $\lambda_L = 2\pi c/\omega_L = 1064$ nm, $L = 25$ mm, $P = 4$ mW, $m = 15$ ng, $\omega_m/(2\pi) = 275$ kHz, the mechanical quality factor $Q = \omega_m/\gamma_m = 2.1 \times 10^4$. Further in the high temperature limit $k_B T \gg \hbar\omega$, we have $\coth(\hbar\omega/2k_B T) \approx 2k_B T/\hbar\omega$.

### A. FROM ROOM TEMPERATURE ($T = 300$ K) TO ABOUT SUB-KELVIN TEMPERATURES.

If we choose $\kappa = 10^8$ s$^{-1}$, $F = 188.4$ s$^{-1}$, $G = 0$, to satisfy the stability conditions (12), the detuning has to satisfy $\Delta_0 \geq 4 \times 10^6$ s$^{-1}$. The Fig. 2 gives the variations of the $\chi q_s$, the effective temperature $T_{\text{eff}}$, and the parameter $r$ with the detuning $\Delta_0$. It should be borne in mind that for the range of the detuning shown in the Fig. 2, $\Delta = \Delta_0 - \chi q_s \approx \Delta_0$. We find the $\chi q_s$ is single valued, so the movable mirror is monostable. Note that the parameter $r$ is very close to unity, $\frac{1}{2}m\omega_m^2 \langle q^2 \rangle \approx \frac{\langle p^2 \rangle}{2m}$; the mirror is thus in nearly thermal equilibrium. The Fig. 2 shows the possibility of cooling the mirror to a temperature of 15.23 K for $\Delta_0 = 4.9 \times 10^7$ s$^{-1}$, which is in agreement with the previous calculation [16].

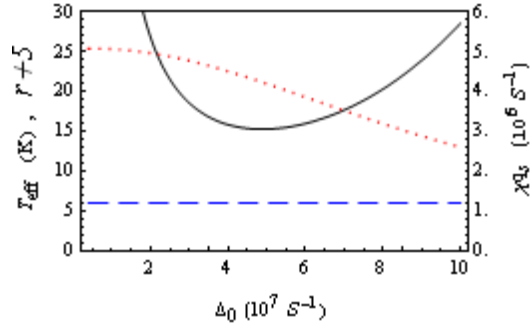

FIG. 2 (color online). The dotted curve indicates the $\chi q_s$ ($10^6 \text{s}^{-1}$) as a function of the detuning $\Delta_0$ ($10^7 \text{s}^{-1}$) (rightmost vertical scale). The solid curve shows the effective temperature $T_{eff}$ (K) as a function of the detuning $\Delta_0$ ($10^7 \text{s}^{-1}$) (leftmost vertical scale). The dashed curve represents the parameter $r$ as a function of the detuning $\Delta_0$ ($10^7 \text{s}^{-1}$) (leftmost vertical scale). Parameters: cavity decay rate $\kappa = 10^8 \text{ s}^{-1}$, cavity finesse $F = 188.4 \text{ s}^{-1}$, parametric gain $G = 0$.

Now we keep the values of $\kappa$, and $F$ the same as in the Fig. 2, and we choose parametric gain $G = 3.5 \times 10^7 \text{ s}^{-1}$ and parametric phase $\theta = 0$, the detuning has to satisfy $\Delta_0 \geq 5.7 \times 10^7 \text{ s}^{-1}$. If $\Delta_0 < 5.7 \times 10^7 \text{ s}^{-1}$ and for fixed $\kappa$ and $G$, the system will be unstable. The threshold for unstable behavior occurs when any of the three conditions (12) is not satisfied. It may be noted that the threshold for parametric oscillation has been of great importance in connection with the production of nonclassical–squeezed light. Near the parametric thresholds but under (13), large degrees of squeezing was produced [18,19]. Thus it would be advantageous to work near the threshold of instability but below the instability point. The Fig. 3 shows the variations of the $\chi q_s$, the effective temperature $T_{eff}$, and the parameter $r$ with the detuning $\Delta_0$. We find the $\chi q_s$ is still single valued, so the movable mirror is still monostable. The minimum temperature reached is $0.65 \text{ K}$ for $\Delta_0 = 6.7 \times 10^7 \text{ s}^{-1}$. Thus with parametric amplifier the minimum temperature is about a factor of 20 lower than the one without parametric interaction. Note that the parameter $r$ is always larger than 1, implying that momentum fluctuations are suppressed over position fluctuations. Note that as one moves away from the threshold for parametric instability, the minimum temperature does not rise sharply which is in contrast to the behavior in the Fig. 2, and is advantageous in giving one flexibility about the choice of the detuning parameter.

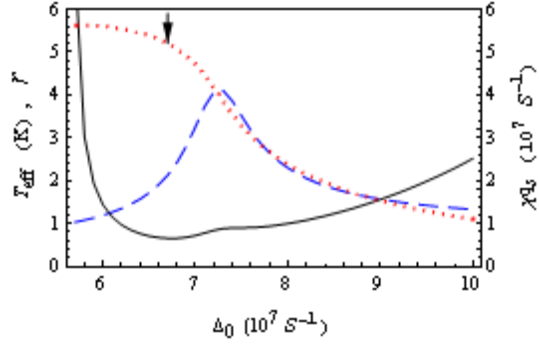

FIG. 3 (color online). The dotted curve indicates the $\chi q_s$ ($10^7 s^{-1}$) as a function of the detuning $\Delta_0$ ($10^7 s^{-1}$) (rightmost vertical scale). The position that corresponds to the minimum effective temperature reached is indicated by the arrow. The solid curve shows the effective temperature $T_{eff}$ (K) as a function of the detuning $\Delta_0$ ($10^7 s^{-1}$) (leftmost vertical scale). The dashed curve represents the parameter $r$ as a function of the detuning $\Delta_0$ ($10^7 s^{-1}$) (leftmost vertical scale). Parameters: cavity decay rate $\kappa = 10^8 \, s^{-1}$, cavity finesse $F = 188.4 \, s^{-1}$, parametric gain $G = 3.5 \times 10^7 \, s^{-1}$, parametric phase $\theta = 0$.

We next examine the case when the behavior of the system is multistable. For this purpose, we choose the cavity to have higher quality factor. We choose $\kappa = 10^7 \, s^{-1}, F = 1884 \, s^{-1}, G = 5 \times 10^6 \, s^{-1}, \theta = 3\pi/4$, then to satisfy the stability conditions (12), the detuning has to satisfy $\Delta_0 \geq 1.847 \times 10^7 \, s^{-1}$. The Fig. 4 gives the behavior of $\chi q_s$ as a function of the detuning $\Delta_0$. We find the $\chi q_s$ is multivalued, so the movable mirror is multistable. By use of the lowest curve of the $\chi q_s$, we obtain the variations of the effective temperature $T_{eff}$ and the parameter $r$ with the detuning $\Delta_0$, as shown in the Fig. 5. We choose the range of the detuning is $2.0 \times 10^7 \, s^{-1} - 3.0 \times 10^7 \, s^{-1}$. The minimum temperature achieved is 0.265 K for $\Delta_0 = 2.0 \times 10^7 \, s^{-1}$. Note that $r$ is close to unity but larger than unity. The general trend is clear. By playing around with various parameters like laser power, cavity finesse, parametric gain, one can achieve a variety of different temperatures. As another example, if we choose $\kappa = 5 \times 10^6 \, s^{-1}$, $F = 3768 \, s^{-1}$, $G = 10^7 \, s^{-1}$, $\theta = 0.2467 + \pi/2$, then we find that the minimum temperature is 0.092 K for $\Delta_0 = 2.13 \times 10^7 \, s^{-1}$.

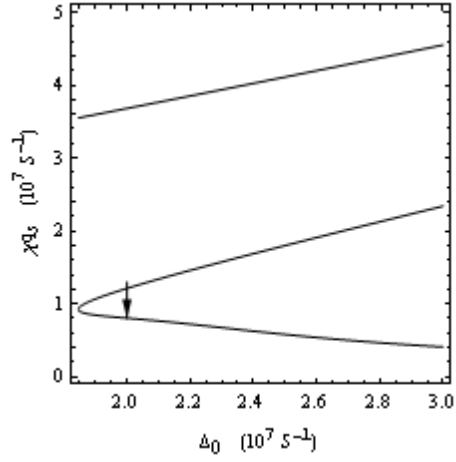

FIG. 4. The behavior of $\chi q_s$ $(10^7 \text{s}^{-1})$ shown as a function of the detuning $\Delta_0$ $(10^7 \text{s}^{-1})$. The position that corresponds to the minimum effective temperature reached is indicated by the arrow. Parameters: cavity decay rate $\kappa = 10^7$ s$^{-1}$, cavity finesse $F = 1884$ s$^{-1}$, parametric gain $G = 5 \times 10^6$ s$^{-1}$, parametric phase $\theta = 3\pi/4$.

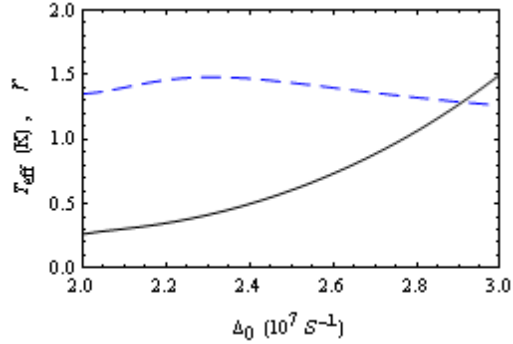

FIG. 5 (color online). The solid curve shows the effective temperature $T_{eff}$ as a function of the detuning $\Delta_0$ $(10^7 \text{s}^{-1})$. The dashed curve represents the parameter $r$ as a function of the detuning $\Delta_0$ $(10^7 \text{s}^{-1})$. Parameters: cavity decay rate $\kappa = 10^7$ s$^{-1}$, cavity finesse $F = 1884$ s$^{-1}$, parametric gain $G = 5 \times 10^6$ s$^{-1}$, parametric phase $\theta = 3\pi/4$.

### B. FROM 1 K TO MILLIKELVIN TEMPERATURES.

If the thermal bath is cryogenically cooled down to a temperature of 1 K and the mirror is initially thermalized, then we can use radiation pressure effects and photon statistics to reach millikelvin or even lower temperatures.

If we choose $\kappa = 10^8$ s$^{-1}$, $F = 188.4$ s$^{-1}$, $G = 0$, the effective temperature $T_{eff}$ with the detuning $\Delta_0$ is shown in the Fig. 6. The minimum temperature reached is $0.051$ K for $\Delta_0 = 4.9 \times 10^7$ s$^{-1}$. Next we examine how the effective temperature changes by the parametric interactions inside the cavity. We keep all other parameters as in the Fig. 6 and choose parametric gain $G = 3.5 \times 10^7$ s$^{-1}$ and phase $\theta = 0$. Then the effective temperature $T_{eff}$ with the detuning $\Delta_0$ exhibits behavior as shown in the Fig. 7. The minimum temperature achieved is $0.0044$ K for $\Delta_0 = 7.9 \times 10^7$ s$^{-1}$, a factor of 12 lower than the one without parametric interaction.

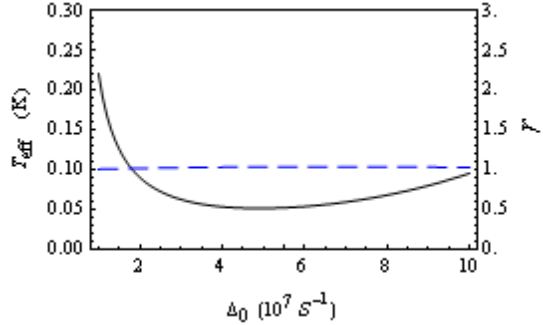

FIG. 6 (color online). The solid curve shows the effective temperature $T_{eff}$ (K) as a function of the detuning $\Delta_0$ ($10^7$s$^{-1}$) (leftmost vertical scale). The dashed curve represents the parameter $r$ as a function of the detuning $\Delta_0$ ($10^7$s$^{-1}$) (rightmost vertical scale). Parameters: cavity decay rate $\kappa = 10^8$ s$^{-1}$, cavity finesse $F = 188.4$ s$^{-1}$, parametric gain $G = 0$.

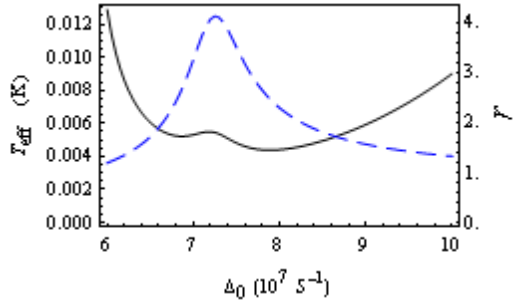

FIG. 7 (color online). The solid curve shows the effective temperature $T_{eff}$ (K) as a function of the detuning $\Delta_0$ ($10^7$s$^{-1}$) (leftmost vertical scale). The dashed curve represents the parameter $r$ as a function of the detuning $\Delta_0$ ($10^7 \text{s}^{-1}$) (rightmost vertical scale). Parameters: cavity decay rate $\kappa = 10^8 \text{ s}^{-1}$, cavity finesse $F = 188.4 \text{ s}^{-1}$, parametric gain $G = 3.5 \times 10^7 \text{ s}^{-1}$, parametric phase $\theta = 0$.

Finally it should be borne in mind that the radiation pressure depends on the number operator and then it is sensitive to the photon statistics of the field in the cavity. The photon statistics can be calculated from the quantum Langevin equations (8). It can be proved that the Wigner function $W$ of the field in the cavity is Gaussian of the form $\exp[\mu(\alpha - c_s)^2 + \nu(\alpha^* - c_s^*)^2 + \lambda(\alpha - c_s)(\alpha^* - c_s^*)]$, with $\mu, \nu, \lambda$ determined by $\kappa, \Delta, G, \theta$ etc. The photon number distribution [24] associated with such a Gaussian Wigner function depends in an important way on the parameter $\mu$ and the inequality of $\mu$ and $\nu$. The latter depend on $G \neq 0$ or on the presence of OPA in cavity.

## V. CONCLUSIONS

In conclusion, we have demonstrated how the addition of a parametric amplifier in a cavity can lead to cooling of the micromirror to a temperature which is much lower than what is achieved in an identical but without parametric amplifier. The parametric processes inside the cavity change the quantum statistics of the field in the cavity. This change leads to lower cooling since the radiation pressure effects are dependent on the photon number. Thus photon statistics becomes central to achieving lower cooling temperatures. The use of parametric process could provide us a way to cool the mirror to its quantum ground state or even squeeze it.

We thank M. S. Kim and M. Paternostro for interesting correspondence on the cooling of the mirror. We gratefully acknowledge support for NSF Grant No.CCF 0829860.

15